\documentclass[aps,prb,twocolumn,showpacs,showkeys,amsmath,amssymb,superscriptaddress]{revtex4-1}
\usepackage[T1]{fontenc}
\usepackage[latin9]{inputenc}
\usepackage{graphicx}

\begin{document}

\title{A simple concentration-dependent pair interaction model for large-scale
simulations of Fe-Cr alloys}

\author{M. Levesque}

\affiliation{CEA, DEN, Service de Recherches de Métallurgie Physique, F-91191
Gif-sur-Yvette, France}

\author{E. Martínez}

\altaffiliation[Correspondent author. Current address: ]{Los Alamos National Laboratory}

\email{enriquem@lanl.gov}

\affiliation{CEA, DEN, Service de Recherches de Métallurgie Physique, F-91191
Gif-sur-Yvette, France}

\author{C-C. Fu}

\affiliation{CEA, DEN, Service de Recherches de Métallurgie Physique, F-91191
Gif-sur-Yvette, France}

\author{M. Nastar}

\affiliation{CEA, DEN, Service de Recherches de Métallurgie Physique, F-91191
Gif-sur-Yvette, France}

\author{F. Soisson}

\affiliation{CEA, DEN, Service de Recherches de Métallurgie Physique, F-91191
Gif-sur-Yvette, France}

\begin{abstract}
This work is motivated by the need for large scale simulations to extract physical information on
the iron-chromium system which is a binary model alloy for ferritic steels
used or proposed in many nuclear applications. From first principle calculations
and the experimental critical temperature we build a new energetic
rigid lattice model based on pair interactions with concentration and temperature
dependence. Density Functional Theory calculations in both norm-conserving and projector augmented wave approaches have been performed. A thorough comparison of these two different \textit{ab initio} techniques leads to a robust parameterization of the Fe-Cr Hamiltonian.
Mean field approximations and Monte Carlo
calculations are then used to account for temperature effects.
The predictions of the model are in agreement with the most recent phase diagram at all
temperatures and compositions. The solubility of Cr in Fe below 700 K remains in the range of about 6\% to 12\%. It reproduces the transition
between the ordering and demixing tendency and the spinodal decomposition limits are also in agreement with the values given in the literature.
\end{abstract}


\maketitle

\section{Introduction\label{sec: Introduction}}

As it has been extensively reported in the literature \cite{Caro2005,Malerba2008,Yvon2009,Klueh2007},
ferritic steels with a content in Cr ranging from 5 to 13 at. \% present
a set of features concerning their radiation damage resistance that
makes them the strongest candidates for future nuclear energy applications as structural materials.
To develop new materials capable of working at high irradiation doses
we need to understand both their thermodynamic and kinetic properties.
At high temperature, the kinetic evolution is rapid enough to observe the formation
of the $\gamma$ phase and the decomposition of Fe-Cr alloys
into two body centered cubic (BCC) solid solutions, $\alpha$
and $\alpha^\prime$ and therefore the phase diagram is well-known. Below 700 K, the equilibrium state is still in
debate. The phase diagrams available in usual compilations and database
like CALPHAD \cite{Saunders-Book,SGTE,Kubaschewski1982} are derived
from high temperature experiments \cite{Williams1974,Hertzman1982,Andersson1987,Chuang1987,Lin1987}.
They display an almost symmetrical $\alpha$-$\alpha^\prime$ miscibility
gap and yield a zero solubility limit of Cr in Fe at low temperature.
However, first principles calculations by Hennion \cite{Hennion1983}
in 1983, confirmed a few years later by a neutron study of short range
order (SRO) \cite{Mirebeau1984,Mirebeau2010} have shown that Fe-Cr
alloys display an ordering tendency for low chromium contents. This
anomaly has been extensively studied using various \textit{ab initio} methods and is now well understood, although the Cr concentration at
which the sign of the mixing energy changes depends on the approach 
(from 5\% up to approximately 10\%) \cite{Olsson2003,Olsson2006,Klaver2006,Paxton2008}.
This behavior has been rationalized in
terms of an anti-alignment of the magnetic moment of Cr in the Fe
matrix, the repulsion between first nearest neighbor (1nn) Cr, and
the ordering tendency observed at low concentrations \cite{Klaver2006}.
Moreover, a few experimental observations in alloys submitted to irradiation
have been recently reviewed by Bonny \emph{et al}. \cite{Bonny2008}.
They suggest that the chromium solubility remains above 8\%, even
at low temperatures. This interpretation is based on the assumption
that irradiation only results in an enhancement of diffusion and that
more complex effects that could modify the solubility limit, such
as ballistic disordering or radiation induced segregation, can be
neglected. Recent critical reviews have therefore highlighted the need to 
modify the Fe-Cr phase diagram at low temperature \cite{Xiong2010,Bonny2008,Bonny2010}. 

Several atomistic models have been proposed in this context to reproduce the complex
thermodynamic behavior of Fe-Cr alloys. Semi-empirical potentials
have been developed that take into account the change of sign of the
mixing energy, such as the concentration dependent model (CDM) of
Caro \textit{et al}. \cite{Caro2005} or the Two-Band Model (2BM)
of Olsson \textit{et al.} \cite{Olsson2005} recently updated by Bonny \textit{et al.} \cite{Bonny2011}. A lot of work has been
done in order to assess their thermodynamic properties as well as
their dynamical behavior \cite{Bonny2009,Erhart2008,Erhart2008-2,Bonny2009-2}.
However, it remains difficult to develop a potential fitting simultaneously
all the key properties that control the thermodynamics and kinetics
of the Fe-Cr decomposition (such as the mixing energies, the point
defects formation energies and migration barriers, with their dependence
on the local atomic distribution and with the corresponding vibrational
entropy contributions). Furthermore, magnetic contributions have not
been introduced in these potentials. Because the kinetic modeling of phase transformations including atomic relaxations and vibrational contributions is a challenging task \cite{Bocquet2002}, these potentials are usually mapped on a rigid lattice model which in turn affects their thermodynamic and kinetic properties. 

Cluster expansion (CE) techniques, based on a
rigid lattice approximation with $N$-body concentration-independent
interactions, have been proposed to model the thermodynamics of Fe-Cr
alloys \cite{Inden2008,Lavrentiev2007,Nguyen-Manh2008}. However,
to be able to reproduce the \textit{ab initio} mixing energies of
Fe-Cr, a purely chemical CE (\textit{i.e.} that does not take explicitly into account the magnetic moments) requires a large set of 
many-body interactions 
\cite{Nguyen-Manh2008,Lavrentiev2007}. The difficulty of obtaining
a small set of effective interactions was also reported by authors
using a screened generalized perturbation method \cite{Ruban2008}.
Therefore, a cluster interaction model, although restricted to a rigid
lattice description, remains quite heavy numerically. The whole corresponding
phase diagram was not published, but following the trend of the empirical
concentration-dependent energy models, it is expected that the critical
temperature for the miscibility gap is way too high compared to experiments
and classical CALPHAD database \cite{Saunders-Book,Andersson1987}.
Two missing ingredients in those CE models are the vibrational entropy
(which is significant in the Fe-Cr system \cite{Swan-Wood2005,Lucas2008})
and the magnetic contributions. The screen generalized perturbation
model has been used to show how the effective cluster interactions
depend on the magnetic state of the alloy and therefore on the temperature
\cite{Ruban2008} and the composition \cite{Korzhavyi2009}. Mixed
models including chemical and magnetic interactions have been proposed
\cite{Ackland2006,Inden2008,Lavrentiev2010,Lavrentiev2010-2}. The Ising model by Ackland, with magnetic moments of constant amplitude,
reproduces some key features of the Fe-Cr alloys. The magnetic cluster
expansion of Lavrentiev \textit{et al.} \cite{Lavrentiev2010,Lavrentiev2010-2}
is able to reproduce the \textit{ab initio} mixing energies with much
fewer interactions than a purely chemical CE and it can take into
account the variation of the magnetic moments with the concentration,
but its phase diagram has never been calculated in the $\alpha$-$\alpha^\prime$
region. The last model we wanted to mention is the Stoner Hamiltonian developed by Nguyen-Manh and Dudarev \cite{Nguyen-Manh2009}.
It is shown that all the significant features of the Fe-Cr alloys
can be explained in terms of bonding effects involving 3\textit{d}
electron orbitals and magnetic symmetry-breaking effects resulting
from intra-atomic on-site Stoner exchange. The complete phase diagram
has not been reported for this Hamiltonian and, as it is said in the
manuscript, further approximate computational algorithms will have
to be developed suitable for large scale simulations. 

Finally, it is worth noting that using a magnetic model in a kinetic
simulation of the $\alpha$-$\alpha^\prime$ decomposition (such
as a kinetic Monte Carlo simulation) would require the relaxation
of the atom-vacancy exchange events and magnetic moments transitions which probably occur
at very different time scale. Even with simplifying assumption (e.g.
if the relaxation time of the magnetic moment is negligible), it would
make the simulation much more time consuming than for a non-magnetic
model. We propose here an alternative model: a concentration and temperature
dependent pair interaction model fitted on \textit{ab initio} calculations
and the experimental critical temperature of Fe-Cr alloys. The goal
is to keep the model simple enough to be used in kinetic Monte Carlo
simulations such as the one of Ref.\onlinecite{Soisson2007}. 

The manuscript
is organized as follows. In Sec.\ref{sec:dft} the results from\textit{
}\textit{\emph{density functional theory}} (DFT) calculations on the
energetics of the Fe-Cr system, that have been used to parameterize
the interaction model, are reported. Because energy values, in the magnetic Fe-Cr system, depend on the method, we have performed our own \textit{ab initio} calculations using two different methods. In Sec.\ref{sec:Thermodynamic-model}
we present the concentration and temperature dependent pair interaction
model and its phase diagram, computed in a mean-field approximation
and by Monte Carlo simulations. The phase diagram, including the spinodal
decomposition region, and the short range order are compared with
available experimental data. Comparison with other models are discussed
in Sec.~\ref{sec:Discussion}. Finally, some conclusions and perspectives
are highlighted.

\section{DFT calculations\label{sec:dft}}

Several studies have already been devoted to the calculations of energetic properties of FeCr 
alloys \cite{Olsson2003,Olsson2006,Klaver2006,Paxton2008,Erhart2008}. 
We have nevertheless performed a new systematic first principles study in order to parameterize our interaction model in a self-consistent way. 
In particular, we have calculated the enthalpy
of mixing of the FeCr alloy to account for its behavior in the whole
concentration range. We have also estimated the interactions between
two Cr(Fe) impurities in a bcc Fe(Cr) matrix, which allows to determine
the cutoff of interactions distance of the pair-interaction model.
Ferromagnetic (FM) Fe and (100)-layered antiferromagnetic (AF) Cr have
been taken as reference states to obtain the values mentioned above.
Note that even though the experimental magnetic ground state of pure
bcc Cr is an incommensurate spin-density wave (SDW), the presence
of Fe atoms seems to reduce the stability
of such a long-ranged state. It, indeed, becomes unstable
against the formation of AF structures with 1.6 $\%$ of Fe \cite{Fawcett_alloy}.
Because our interest is mainly focused on the Fe-rich side of the
alloy, we assume the AF state for Cr in the present study.

Calculations are performed in the framework of Density Functional
Theory as implemented in the PWSCF code \cite{PWscf}. They are spin
polarized within the Generalized Gradient approximation (GGA) with the PBE parametrization \cite{Perdew1996}. We have used the Projector
Augmented Wave (PAW) potential instead of pseudopotentials. The kinetic
energy cutoff chosen for the plane-wave basis set was 544 eV. All the calculations are fully relaxed, i.e., both atomic positions
and simulation-cell volumes are optimized. The corresponding residual
force and stress tolerances are respectively 0.04 eV/{\AA} and 5
kbar. We have also calculated the mixing enthalpies using norm-conserving (NC) pseudopotentials
and localized basis sets, as implemented in the SIESTA code \cite{siesta}.
This approach has been shown to give results of equivalent accuracy
as plane-wave DFT methods. In particular, properties
of defects in various Fe based systems have been satisfactorly predicted
\cite{Fu2004,Fu2005,Fu2005_2,Soisson2007}. It is however less computationally
demanding thanks to significant reduction of the basis size. The aim
is to check the ability of this less standard DFT approach for quantitative
prediction of properties in FeCr alloys, where the energetics may
be extremely sensitive to magnetic couplings \cite{Klaver2006,Soulairol2010}.

\subsection{Mixing Enthalpy}

\label{subsec: Hmix} 
The enthalpy of mixing is defined as: 
\begin{equation}
\Delta H_{mix}=\frac{E[n\mathrm{Fe}+m\mathrm{Cr}]-\{nE[\mathrm{Fe}]+mE[\mathrm{Cr}]\}}{n+m}
\end{equation}
where $E{\rm [\mathit{n}Fe+\mathit{m}Cr]}$ is the total energy of
a mixed system containing $n$ Fe atoms and $m$ Cr atoms. $E(\mathrm{Fe})$
and $E(\mathrm{Cr})$ are energies per atom of the Fe and
Cr reference systems. Because the calculations are performed at zero
pressure this value is also equivalent to the mixing energy.

The supercells used for different concentrations have been generated
using two methods. The first generation method is \textquotedbl{}user\textquotedbl{}-chosen.
It consists of a large set of ordered structures devised to explore various energetic landscapes ($\mathrm{DO_{3}}$,
B2, Fe$_{n-1}$Cr, FeCr$_{m-1}$ etc.), the same as considered in
a previous work on Fe-Cu alloys \cite{Soisson2007}. The other generation method is
based on the special quasi-random structure methodology (SQS) \cite{Zunger1990})
which allows to generate a supercell with as small short
range order as possible. These supercells are thus the best representative configurations of
a random solid solution for each concentration. The number of atoms
in each supercell is either 54 or 128.

\begin{figure}
\includegraphics[width=8.5cm]{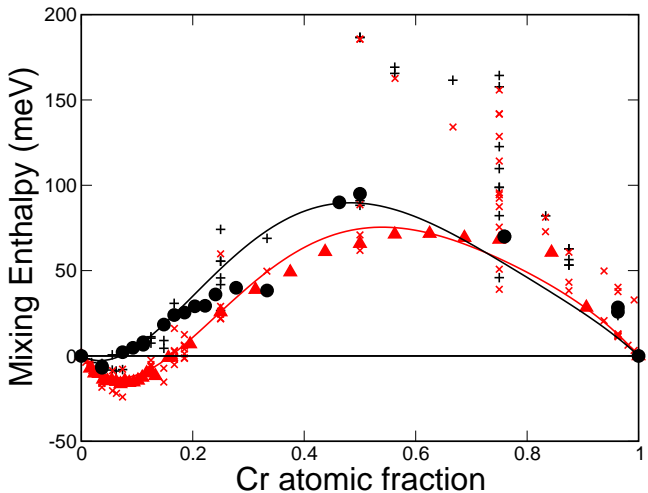}

\caption{\label{fig:Hmix} Enthalpy of mixing for the Fe-Cr system as a function
of Cr atomic fraction. PAW-GGA (NC-GGA) calculations are in
black (red). Full symbols are for SQS structures and crosses for ordered structures.
The lines give the fit of the enthalpy of mixing of the SQS by the
Redlich-Kister formula (see section\,\ref{sec:Thermodynamic-model}).}
\end{figure}

The resulting mixing enthalpies for both ordered and SQS structures
are shown in Fig. \ref{fig:Hmix}.

In good agreement with previous DFT results \cite{Olsson2007,Klaver2006},
we note a change of sign of the mixing enthalpy, showing
negative values for low Cr concentrations according to both PWSCF
and SIESTA approaches. However, the range of this negative part of
the enthalpy of mixing as well as its depth strongly depend on the
approach. From the SIESTA-NC calculations the change in sign is around
$x_{Cr}=0.15$ while with the PWSCF-PAW approach, the value is about
$x_{Cr}=0.07$.

First of all, we focus on two extreme cases, i.e., infinite dilution
in Fe and Cr respectively, where we may also define the solution energy
as: 
\begin{equation}
E_{sol}^{XinY}=E[\mathrm{Y}_{n-1}\mathrm{X}_{1}]-\{(n-1)E[\mathrm{Y}]+E[\mathrm{X}]\}
\end{equation}
where X and Y are either Fe or Cr, $E[\mathrm{Y}_{n-1}\mathrm{X}_{1}]$
is the total energy of a supercell containing \textit{n-1} atoms of Y and one
atom of X and $E[Y]$ and $E[X]$ are the energy per atom of the pure
systems: bcc FM Fe and AF Cr. The solution energies
are well converged within 1meV for $n=128$ atoms. For the case of Cr dissolution
in Fe we find a value of $E_{sol}=-0.20$ eV using PWSCF while SIESTA
predicts $E_{sol}=-0.47$ eV. They are consistent with previous
DFT values ranging between $E_{sol}=-0.12$ eV and $E_{sol}=-0.46$
eV \cite{Klaver2006,Olsson2007}. We see that all DFT results predict
Cr dissolution to be exothermic, \textit{i.e.}, energeticaly favorable
to insert one substitutional Cr in the Fe matrix. However, the precise
value reveals to be method dependent. In particular, the SIESTA-NC result
overestimates the solution tendency of Cr in Fe with respect to the
PWSCF-PAW data. In order to gain more insight into the origin of this
overstimation, we have performed complementary PWSCF calculations
using a NC pseudopotential (PWSCF-NC) as close as possible to that
of SIESTA. The obtained Cr solution energy in Fe is -0.49 eV, very
close to the SIESTA value. This comparison suggests that the overestimation
of Cr solution energy is essentially due to the NC-pseudopotential
approximation rather than the use of localized basis functions in
the SIESTA approach. We have also checked that the magnitude of Cr
solution energy is indeed closely correlated with the local magnetic
moments of the Cr in Fe. The corresponding values from the PWSCF-PAW,
PWSCF-NC and SIESTA-NC studies are 2.1 $\mu_{B}$, 2.6 $\mu_{B}$
and 2.5 $\mu_{B}$ respectively. It is interesting to point out that
the overstimation of the Cr solution energy from a NC pseudopotential
prediction is closely correlated to the obtained higher value of Cr
local moments with respect to the PAW value.

When a Cr atom is substituted by one Fe atom in the Cr matrix, the
solution energy obtained was $E_{sol}=0.45$ eV using PWSCF and $E_{sol}=0.29$ eV with SIESTA, 
indicating an endothermic reaction. In this
case, the Fe local magnetic moment found for the Fe impurity is within the precision limits, 0.02 $\mu_{B}$ while SIESTA gives 0.14 $\mu_{B}$. In both cases the magnetic moment of the Fe solute is anti-alligned to the local moment of its first
nearest neighbors. The local magnetic moment of all the Cr atoms remains,
as expected, practically the same as in pure AF Cr. The small moment
of Fe may be explained as a consequence of magnetic frustration resulting
from the competition between the Fe and its first- and second-nearest
Cr neighbors. As also suggested by a previous study \cite{Klaver2006}, 
Fe and Cr first and second nearest neighbors prefer an antiferromagnetic coupling, which
can clearly not be satisfied for an isolated Fe in a bcc AF Cr lattice.

Beyond the infinite diluted cases, ordered structures with a mixing energy lower than
the SQS-random configurations ($\rm \Delta H_{mix} = -x E_{sol}^{Cr in Fe}$) are observed at low Cr concentrations with both PWSCF
and SIESTA. In particular,
the $\mathrm{Fe}_{52}\mathrm{Cr_{2}}$ system with the 2 Cr atoms
separated by (1.5,1.5,1.5) times the bcc lattice parameter (a$_{0}$)
has an energy lower than the solid solution of the same composition,
suggesting the possible formation of an intermetallic phase for that concentration at low temperatures. Indeed, the same $\mathrm{Fe}_{52}\mathrm{Cr}_{2}$
structure has also been pointed out by Erhart \textit{et al}. as a
possible intermetallic system \cite{Erhart2008}. Other DFT calculations predict that the Fe$_{15}$Cr \cite{Lavrentiev2007} or the Fe$_{14}$Cr 
\cite{Pareige2009} ordered structures could be the ones forming the intermetallic compounds.
However, it should be noticed that the relative stability of such phases is difficult to assess because 
their difference in formation energies is very close to the DFT uncertainties and because it remains to be verified 
whether they may exist at finite temperatures when the entropy becomes relevant.

On the other hand, at higher Cr concentrations, SQS systems show overall
lower energy than the ordered configurations. This is indeed consistent
with the positive mixing enthalpies suggesting a tendency to phase
separation rather than ordering. 

\subsection{Impurity Interactions}

\label{subsec: imp_int}
\begin{figure}
\centering
\includegraphics[width=8.5cm]{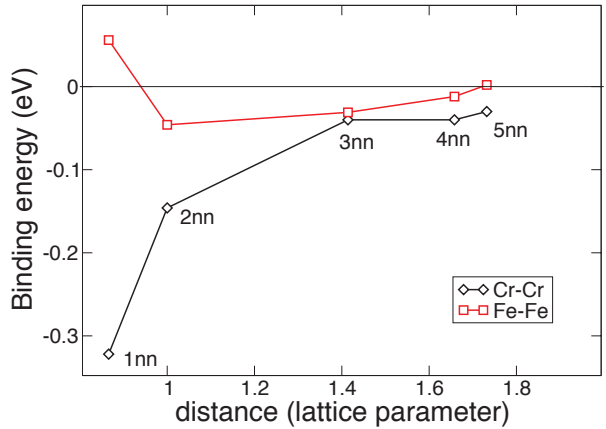}

\caption{Binding energy of two Cr (Fe) impurities in an Fe
(Cr) matrix where $i$nn stands for the $i^{th}$ nearest neighbor
between the impurities in a bcc lattice. }
\label{fig:bind}
\end{figure}

In order to determine the cutoff distance
of the pair interaction model for FeCr, we have also evaluated the
interaction between two Cr(Fe) impurities in a bcc Fe(Cr) matrix.
Binding energy between two X atoms $i^{th}$ nearest neighbors in
a bcc lattice of Y atoms is defined as follows, where positive values
mean attraction: 
\begin{equation}
E^{b}(\mathrm{X}-\mathrm{X})=-E[\mathrm{Y_{N-2}}+\mathrm{X}_{2}]-E[\mathrm{Y}_{N}]+2E[\mathrm{Y}_{N-1}\mathrm{X}_{1}]
\end{equation}
where $E[\mathrm{Y}_{N-2}+\mathrm{X}_{2}]$ is the total energy of
the system with the two X atoms at a $i^{th}$ nearest neighbor distance,
$E[\mathrm{Y}_{N}]$ is the total energy of $N$ Y-atoms in the corresponding
reference system (either ferromagnetic bcc Fe or antiferromagnetic
bcc Cr), and $E[\mathrm{Y}_{N-1}\mathrm{X}_{1}]$ is the total energy
of the system of $N$ atoms with just one impurity atom. The values
for the binding energies of Cr-Cr in Fe and Fe-Fe in Cr are shown in Fig. \ref{fig:bind}. The
calculations have been done within the more accurate PAW approach
using 128-atom supercells.

Consistent with previous DFT calculations \cite{Olsson2007} and with the experimentally observed 
ordering tendency at low Cr content, we find
that two Cr atoms repel each other in a dilute FeCr alloy. Such repulsion
is particularly strong for $1nn$ and $2nn$ interactions. The binding
energies are -0.32 eV and -0.15 eV respectively from the PWSCF calculations.
For $3nn$ to $5nn$ the Cr repulsion becomes significantly weaker
(around -0.04 eV) according to our results  (Fig. \ref{fig:bind}). It vanishes for farther
Cr-Cr distances within the estimated error bar of $\pm$ 0.025 eV.
As explained in previous studies \cite{Klaver2006}, this Cr-Cr repulsion
is directly correlated to the corresponding local magnetic structure.
Local magnetic moments of both Cr atoms are found to be parallel to
each other when they are close neighbors. Also, their moment amplitudes
are reduced as compared with that of an isolated Cr (2.2 $\mu_{B}$).
For instance, we find local moment reductions of around 0.1 $\mu_{B}$
for two $1nn$ and $2nn$ Cr atoms with respect to an isolated Cr.
This can be understood as a magnetic frustration resulting from competition
between various magnetic coupling tendencies, i.e., antiferromagnetic
for Fe-Cr and Cr-Cr and ferromagnetic for Fe-Fe pairs. Indeed, when
performing complementary calculations constraining all the Cr local
moments to zero, the resulting Cr-Cr binding energies become negligible.

On the other hand, in the case of two Fe impurities in a Cr matrix,
their binding energy is slightly positive for a $1nn$ separation (0.06 eV),
whereas it is negative for the $2nn$ Fe-Fe pair (-0.05 eV). Beyond, all the values
are repulsive, but their magnitudes are smaller than 0.03 eV, close to our estimated error bar (Fig. \ref{fig:bind}). It is interesting to mention that the change of interaction
between the $1nn$ to the $2nn$ separations
, \textit{i.e.}, from an attraction to a repulsion, may be linked to a change of local magnetic moments of the respective Fe atoms.
Indeed, as discussed in Sec. \ref{subsec: Hmix}, an isolated Fe in the AF-Cr shows a small moment due to 
the magnetic frustration. It is also the case for all the Fe atoms separated by a $2nn$ distance or farther.
However, the magnetic state can be expected to change when Fe atoms get close to each other. For instance,
when they are first nearest neighbors, one of the two Fe atoms adopts a high moment of 2.11$\mu_{B}$ whereas the other
remains at a low-moment state (0.53$\mu_{B}$). Both Fe moments align parallel to each other, but only the high-moment
Fe is antiferromagnetically coupled to its Cr first-nearest neighbors. This assymetric configuration suggests that at least 
the magnetic frustration of one iron atom, i.e., the high-moment Fe, is partly relaxed, inducing a decrease of the system energy.
It is worth mentioning that other metastable states may also exist for the $1nn$ Fe-Fe case. For instance, we have found another magnetic
configuration where both Fe atoms have low local moments. The corresponding binding energies is practically zero.

Even though the absolute values of Cr-Cr interaction energies in Fe are overall larger than the corresponding Fe-Fe values,
in both cases, the range of the significant interactions
is up to a second-nearest neighbor distance, which may therefore be
reasonably considered as the cutoff distance for our pairwise energetic
model as described below (Sec. \ref{sec:Thermodynamic-model}).

\section{Thermodynamic model\label{sec:Thermodynamic-model}}

\subsection{Constant pair interactions (Ising model)}

Our objective is now to build an interaction model able to take into account the key energetic properties revealed by DFT calculations and to predict a phase diagram in agreement with the experimental one. The most simple model of phase separation in a binary A-B alloy is
probably the Ising model, with constant pair interactions $\epsilon_{AA}^{(i)}$,
$\epsilon_{AB}^{(i)}$ and $\epsilon_{BB}^{(i)}$ between A and B
atoms on $i^{th}$ neighbor sites. The mixing free enthalpy of a solid
solution can be computed using mean-field (MF) approximations (see e.g.\,\cite{Porter1992}).
With the simplest Bragg-Williams (BW) approximation and when $\epsilon_{AA}^{(i)}+\epsilon_{BB}^{(i)}-2\epsilon_{AB}^{(i)}<0$, 
one gets for the mixing enthalpy: 

\begin{equation}
\Delta H_{mix}=-\Omega\: x(1-x)\label{eq:enthalpy_mix}\end{equation}

\noindent While the configurational entropy of mixing is given by:

\begin{equation}
\Delta S_{mix}=-k_{B}\left[\left(1-x\right)\ln\left(1-x\right)+x\ln x\right]\end{equation}

\noindent where $x$ is the B atomic fraction and $k_{B}$ the Boltzmann
constant,

\begin{equation}
\Omega=\sum_{i}\left[\frac{z^{(i)}}{2}\left(\epsilon_{AA}^{(i)}+\epsilon_{BB}^{(i)}-2\epsilon_{AB}^{(i)}\right)\right]
\end{equation}

\noindent is the ordering energy and $z^{(i)}$ the coordination number of shell \textit{i}.
The minimization of the free enthalpy $\Delta G_{mix}=\Delta H_{mix}-T\Delta S_{mix}$
gives a symmetrical miscibility gap, with a critical temperature $T_{c}=-\Omega/2k_{B}$.
In the BW approximation, when all the combination $\epsilon_{AA}^{(i)}+\epsilon_{BB}^{(i)}-2\epsilon_{AB}^{(i)}$
are negative, the phase diagram depends exclusively on $\Omega$ and
not on the distribution of the interactions among the different coordination
shells $(i)$. This approach neglects the short range order in the solid solution. In the specific case of alloys with nearest neighbors interactions, the BW critical temperature is 20\% larger than the Monte Carlo reference value \cite{Domb1974}. The
discrepancy decreases with the range of interactions (for infinitely
long-range interactions, mean-field approximations become exact\,\cite{Domb1966}).

\subsection{Composition-dependent pair interactions}

A constant pair interaction model always gives symmetrical mixing
energies and phase diagrams and therefore cannot reproduce the DFT
mixing energies of Fe-Cr alloys (Fig. \ref{fig:Hmix}), with negative mixing
energies in the Fe-rich configurations only. To be able to reproduce
the mixing enthalpy in the whole concentration range we introduce
pair interactions that depend on the local composition, using a polynomial
expression. In the BW approximation, the mixing enthalpy is 
given by: 
\begin{equation}
\Delta H_{mix}=-\Omega(x)\: x(1-x)=-x(1-x)\sum_{p=0}^{n}{L^{(p)}(1-2x)^{p}}
\end{equation}

\noindent also known as the Redlich-Kister formalism \cite{Redlich-Kister1948}.
$n$ is the maximum order of the parametrization and $L^{(p)}$ is
called interaction parameter of order $p$ and it has the form: 
\begin{equation}
L^{(p)}=a^{(p)}+b^{(p)}T
\end{equation}

The $L^{(p)}$ parameters at 0\,K (i.e. the $a^{(p)}$ parameters)
are fitted on the mixing energies of the SQS structures presented in sec.\,\ref{sec:dft}.
Ordered structures are not taken into account because the SQS configurations are more representative of a random solid solution described by the BW approximation. The best fit we have found (see Fig. \ref{fig:Hmix}) is
of the form: 
\begin{equation}
\Omega(x)=(x-\alpha)(\beta x^{2}+\gamma x+\delta)\label{eq:ordering}
\end{equation}
where the values of $\alpha$, $\beta$, $\gamma$ and $\delta$
for the PWSCF-PAW and SIESTA-NC results are given in table \ref{tab:om_param}. The maximum
of the $\Delta H_{mix}$ in the PWSCF-PAW fit is at $x=0.48$ with
a value of 0.089 eV, whereas for the SIESTA-NC $x=0.52$ and the value
is 0.071 eV. 

\begin{table}[h]
\begin{tabular}{|c|c|c|}
\hline 
 & PWSCF-PAW  & SIESTA-NC \\
\hline 
$\alpha$  & $0.070$  & $0.160$ \\
$\beta (eV)$  & $-2.288$  & $-2.348$ \\
$\gamma (eV)$  & $4.439$  & $4.381$ \\
$\delta (eV)$  & $-2.480$  & $-2.480$ \\
\hline
\end{tabular}

\caption{\label{tab:om_param} Fitting parameters for the ordering energy $\Omega(x)$
obtained from PWSCF-PAW and SIESTA-NC calculations}
\end{table}

\begin{figure}
\includegraphics[width=8.5cm]{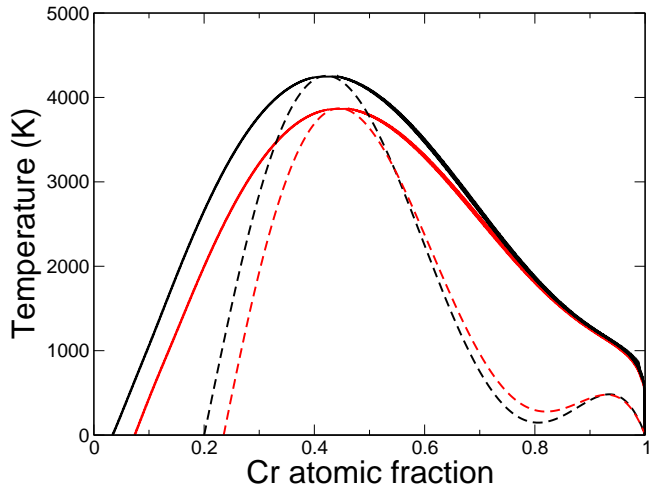}

\caption{\label{fig:pd_cd} Mean field Fe-Cr phase diagram with the concentration-dependent
interaction model (no temperature dependence) fitted on the \textit{ab
initio} PAW (solid black lines) and NC (solid red lines) calculations.
Spinodal decomposition limits are shown in dashed lines.}
\end{figure}

The corresponding phase diagram has been first computed in the BW approximation,
 with $b^{(p)}=0$, i.e. with temperature independent pair interactions 
(see Fig.\ref{fig:pd_cd}). We observe that the solubility limits are
non-symmetric. The solubility of Cr in Fe does not vanish at 0K. On
the other hand, the Fe solubility in Cr is negligible at that temperature.
The non-zero solubility limit on the Fe rich side is in contradiction
with the reference phase diagram given by CALPHAD\,\cite{Andersson1987}
but in accord with more recent studies \cite{Bonny2008,Xiong2010,Bonny2010,Bonny2011}. The
critical temperature (about 4200 K for the PWSCF-PAW parametrization
and 3800 K for the SIESTA-NC values) is much higher than the experimental
one (approximately 1000 \,K \cite{Xiong2010}). The spinodal limits
for both parametrizations, defined as: 
\begin{equation}
\frac{\partial ^2 \Delta G _{mix}}{\partial ^2 x}=0
\end{equation}
are shown in the same Fig. \,\ref{fig:pd_cd}.
We observe an unusual local minimum (for $x \approx 0.8$) on the Cr rich side. This phenomenon
occurs at temperatures lower than 500 K, regime where data is difficult
to obtain experimentally due to slow kinetics. The phase diagram obtained
with the PWSCF-PAW and SIESTA-NC parameters are qualitatively similar.
Since the PWSCF-PAW is more reliable and for the sake of clarity,
we will only discuss the corresponding results in the following.

\begin{figure}

\includegraphics[width=8.5cm]{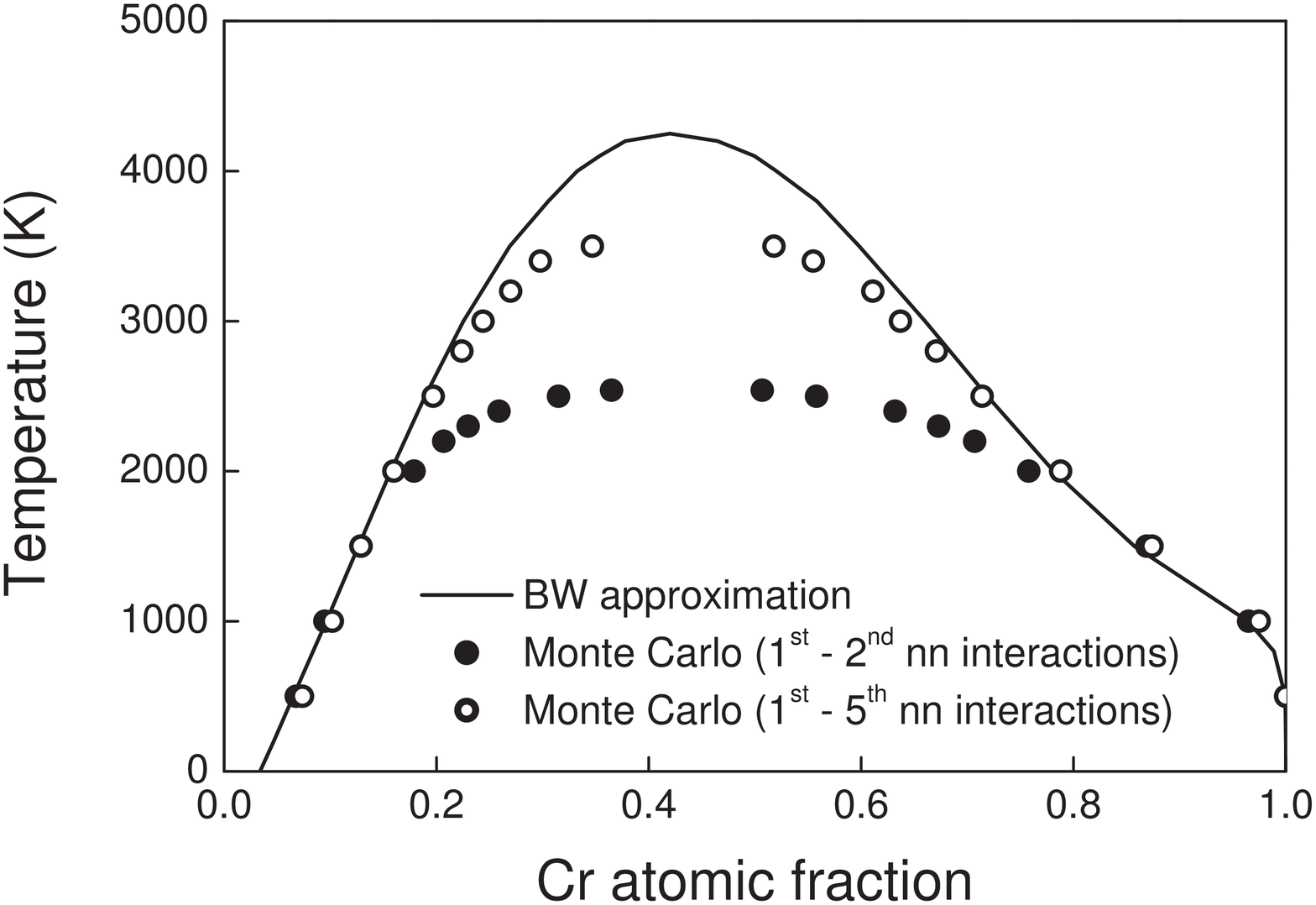}\caption{The Fe-Cr phase diagram with a composition dependent pair interaction
model fitted on the PWSCF-PAW mixing energies (no temperature dependence).
The continuous line gives the solubility limit computed in the BW
approximation. The open circles give the Monte Carlo results with
first and second neighbor interactions, the full circles the Monte
Carlo results with interactions up to the fifth neighbors. \label{fig:the-Fe-Cr-phase_Tind}}

\end{figure}

The phase diagram has also been computed by Monte Carlo simulations
in the semi-grand canonical ensemble \cite{Frenkel2002}. In the BW
approximation, as for the constant interaction model, the phase diagram
only depends on the ordering enthalpy (Eq.\,\ref{eq:ordering}).
For the Monte Carlo simulations for the same ordering energy, one must consider the pair interactions
$\epsilon_{FeFe}^{(i)}$, $\epsilon_{CrCr}^{(i)}$ and $\epsilon_{FeCr}^{(i)}$,
the range of interactions and the way they decrease with the distance.
For the sake of simplicity we have chosen that cross interactions
$\epsilon_{FeCr}^{(i)}$ carry the dependency on the local concentration. 
The self-interactions $\epsilon_{FeFe}^{(i)}$
and $\epsilon_{CrCr}^{(i)}$ are considered as constants given by
the cohesive energies of the pure elements, according to $E_{coh}(A)=-\sum_{i}z^{(i)}\epsilon_{AA}^{(i)}$
(Caro and coworkers followed the same strategy in the development
of their CDM model \cite{Caro2005}). 
The local Cr concentration around a Fe-Cr pair is defined as the
fraction of Cr atoms among their neighbors. If the interactions are
limited to the $r^{th}$ nearest-neighbors, the local chromium concentration
around a Fe atom on site $i$ and a Cr atom on site $j$ is defined
as: 

\[
c(\mathrm{Fe}_{i}\mathrm{Cr}_{j})=\frac{\sum_{n=0}^{r}\sum_{k=1}^{z^{(n)}}p_{ik}^{(n)}+\sum_{n=0}^{r}\sum_{k=1}^{z^{(n)}}p_{jk}^{(n)}}{2\sum_{n=0}^{r}z^{(n)}}\]

\noindent where $p_{ik}^{(n)}=1$ when the $k^{th}$ neighbor of the
site $i$ at a $n^{th}$ neighbor position is occupied by a Cr atom. We also include in the calculation the type of the atoms in sites \textit{i} and \textit{j}.

To assess the effect of the interaction range, we have used two sets
of pair interactions. One has to consider enough neighbors to get
a sufficient discretization of the mixing energy with its change of
sign at 7\%, so that first nearest-neighbor interactions are not enough.
Therefore, the first set of parameters is limited to first and second neighbor
interactions, with the second neighbor interactions two times smaller
than the first ones ($\epsilon_{XY}^{(2)}=\epsilon_{XY}^{(1)}/2$.
The second set includes up to the fifth nearest-neighbors interactions
and they decrease more slowly, as the inverse of interatomic distance. 

The resulting phase diagrams are compared with the BW approximation
on Fig. \,\ref{fig:the-Fe-Cr-phase_Tind}: at low temperature, the BW approximation
is close to the Monte Carlo results. At high temperature it underestimates
the mutual solubility of Fe and Cr and overestimates the critical
temperature, by approximately 40\% when the interactions are limited
to the first and second neighbor shells. The discrepancy is then two
times larger than for the usual Ising model. The critical temperature
of the Monte Carlo simulations is significantly higher with interactions
up to the fifth neighbors. This is in agreement with the usual tendency, where the mean field and Monte Carlo results converge for infinite interaction range \cite{Domb1966}. 

\subsection{Temperature dependence }

The critical temperatures calculated by the composition-dependent approach, shown in Figs. \ref{fig:pd_cd} and \ref{fig:the-Fe-Cr-phase_Tind}, lay well above the critical temperature observed
experimentally for this system, of about 1000 K \cite{Xiong2010}. The CDM
potential shows the same deviation, as does the chemical CE (see section\,\ref{sec:Discussion}). 

We rationalize this difference in terms of the mayor effects that
are not taken into account: 

- The competition between magnetic and chemical interactions.

- The intrinsic nature and the amplitude of the atomic magnetic moments change. The magnetic moments decrease with the temperature
what in turn decreases the pair interaction strength \cite{Fawcett1994}.

- The vibrational entropy. 

- The magnetic entropy. 

\begin{figure}
\includegraphics[width=8.5cm]{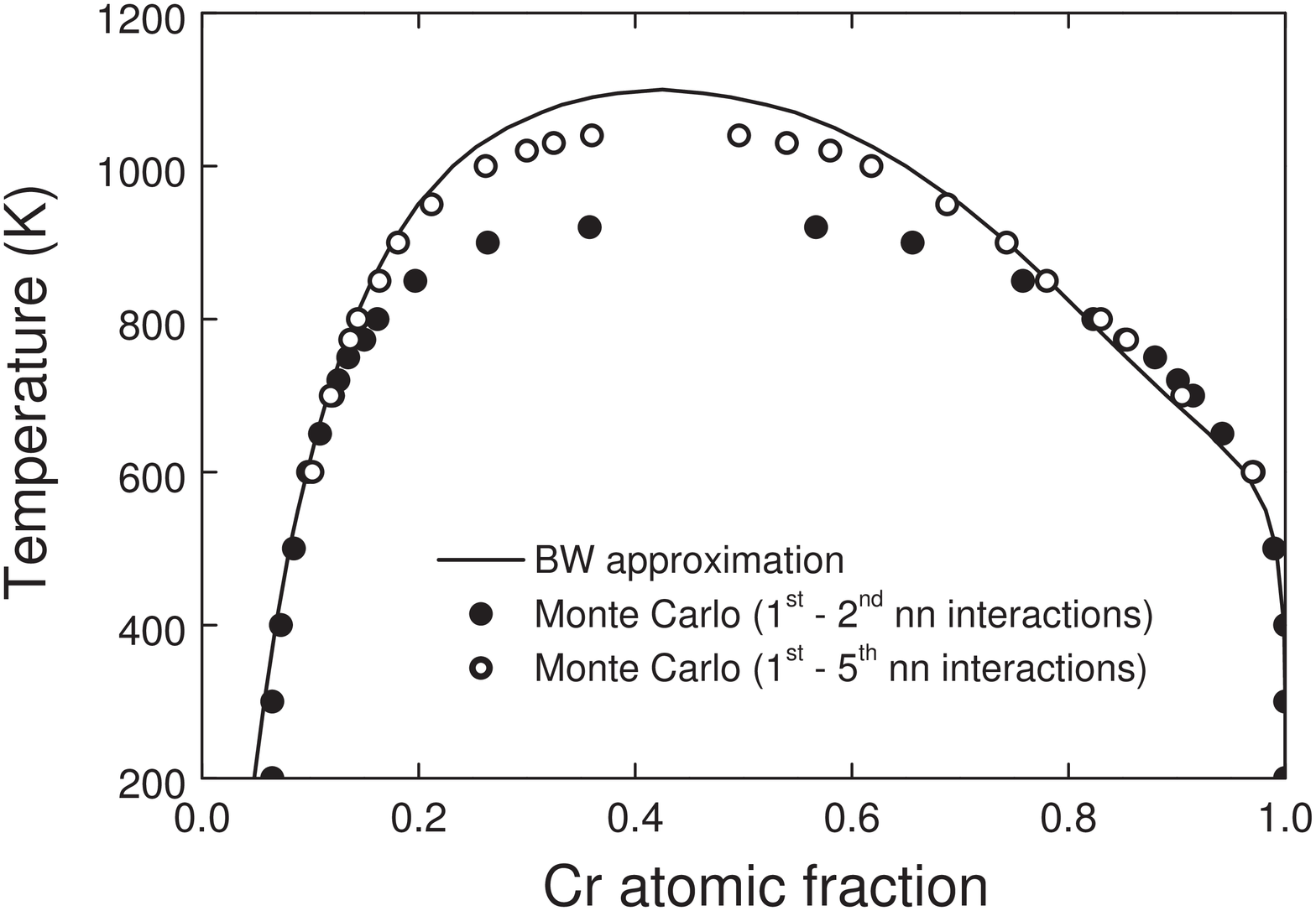}\caption{\label{fig:pd_cdtd} The Fe-Cr phase diagram with a composition and
temperature dependent pair interaction model fitted on the PWSCF-PAW
mixing energies and the experimental critical temperature. The continuous
line gives the solubility limit computed in the BW approximation.
The full circles give the Monte Carlo results with first and second
neighbor interactions, the open circles the Monte Carlo results with
interactions up to the fifth nearest neighbors.}
\end{figure}

\begin{figure}
\includegraphics[width=8.5cm]{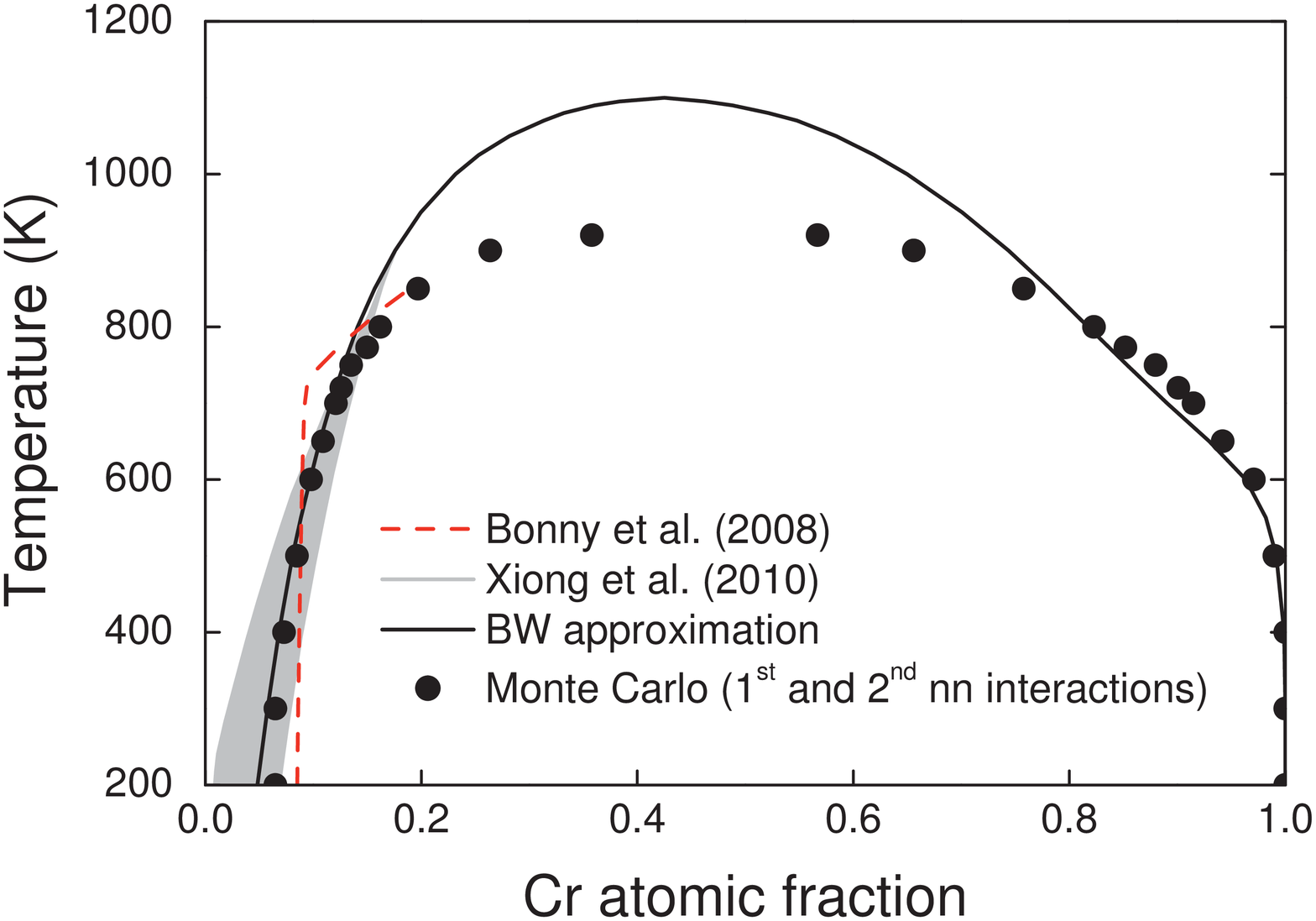}

\caption{\label{fig:PD-model/compil}The Fe-Cr phase diagram. Comparison between
the composition and temperature dependent pair interaction model (Monte
Carlo simulations with first and second neighbor interactions and
BW approximations fitted on PWSCF-PAW mixing energies) and the critical
reviews of Bonny \textit{et al}. \cite{Bonny2008} (dotted line) and
Xiong \textit{et al.} \cite{Xiong2010} (shaded region).}
\end{figure}

One could in principle evaluate the vibrational entropy from DFT calculations, for instance
in the harmonic approximation, but would be obliged to take the rest
of the temperature effects empirically. We consequently decided to
introduce an empirical temperature dependency on the ordering energy
to compensate all the effects of the non-configurational entropies
and magnetic contributions. In order to keep the simplicity of the model and to get a phase diagram closer to the experimental one 
we assign to the ordering energy $\Omega$ a simple linear
dependency on temperature: 
\begin{equation}
\Omega(x,T)=\Omega(x)\left(1-\frac{T}{\Theta}\right)
\end{equation}
where $\Theta$ has units of temperature and it is adjusted such that the Monte Carlo simulations yield the experimental critical temperature ($\approx$ 1000 K). This effect is again taken into account solely to fit the pair interaction cross term $\epsilon_{FeCr}^{(i)}$.
We find $\Theta=1480$ K using the PWSCF-PAW
parameters. The phase diagram as given by this model is shown
in Fig.\,\ref{fig:pd_cdtd}. It is worth noting that it
is not just a temperature rescaling of the one in Fig. \,\ref{fig:the-Fe-Cr-phase_Tind},
since the configurational entropy is not changed. As a consequence
the BW and the Monte Carlo results are closer than before. These results match quite well the collection of experimental
results reported by Xiong \textit{et al}. \cite{Xiong2010} and Bonny
\textit{et al}. \cite{Bonny2008} (Fig.\,\ref{fig:PD-model/compil}).
The Cr solubility in Fe is larger than 0 at low temperatures. Following
the results by Xiong \textit{et al}. the solubility limit at 250
K should be between 1 and 7\%. We have obtained a value of around 7\% with
our model. The solubility limit in the Cr rich side is lower than
1\% at temperatures below 600 K. At very low temperatures the Monte
Carlo results show a solubility of Cr in Fe of about 6\%.

Long runs of Monte Carlo simulations at lower temperatures seem to
confirm that the Cr solubility in Fe at zero temperature is different
from zero. This unusual aspect of a demixing alloy can be explained
by the change of sign of the mixing energy, as already mentioned. No evidence of long-range ordered structure  
has been observed above 200 K. At lower temperature, the stability of such structures is difficult to study because 
the efficiency of the Monte Carlo algorithm decreases. A negative ordering energy
invariant with respect to the local concentration would imply the
stabilization of an ordered phase. At 0 K, the solid solution would
be less energetically favorable than the two-phase system formed
by the ordered phase and a pure phase. However, in the case of a concentration
dependent model, it is possible that in the small concentration
range associated with a negative value of the ordering energy there is no formation
of an ordered phase and then stabilization of the solid solution.

\begin{figure}

\includegraphics[width=8.5cm]{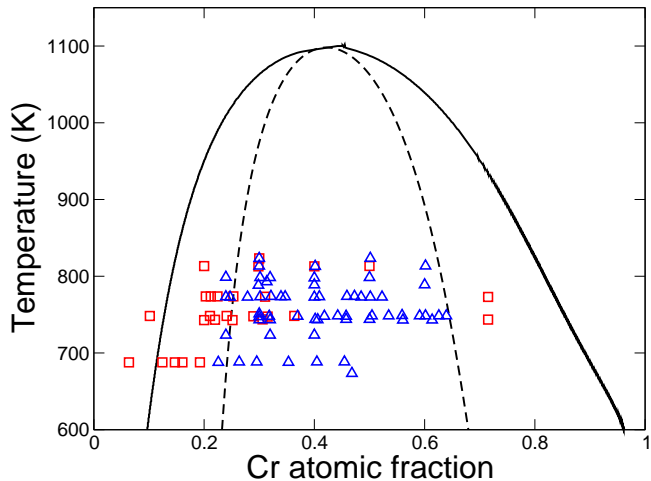}\caption{\label{fig:sp_cdtd}Miscibility gap and spinodal limit of the
composition and temperature dependent pair interaction model (PWSCF-PAW
parameters) computed in the BW approximation. The experimental data for the nucleation and growth regime
are given by square red dots and the spinodal decomposition regime
is given by the blue triangles. The experimental values have been
collected by Xiong \textit{et al}. \cite{Xiong2010}.}

\end{figure}

The spinodal decomposition limits as given by our concentration and
temperature dependent model are shown in Fig. \,\ref{fig:sp_cdtd}
(computed in the BW approximation) where they are compared to the
experimental data compiled by Xiong \textit{et al.} \cite{Xiong2010} obtained in the temperature range
of 650 to 800 K. The existence of a strictly defined limit between two kinetic regimes (nucleation and growth and spinodal decomposition) is debatable \cite{Binder2001}. Nevertheless,
we observe in Fig. \,\ref{fig:sp_cdtd} that the spinodal limits we
are proposing are in good agreement with the existing experimental
data.

\subsection{Short Range order}

\label{subsec:sro} The negative part of the enthalpy of mixing at
low Cr concentrations induces the formation of short range order (SRO)
structures in the Fe-Cr alloy, as shown experimentally
by Mirebeau \textit{et al}. \cite{Mirebeau1984,Mirebeau2010} who
measured the Cowley-Warren SRO parameter for different Cr contents
via neutron diffraction at 703 K. They observed a change in sign in the parameter
at around 10\% Cr, showing a minimum close to 5\%. This inversion
of sign was earlier predicted by Hennion \cite{Hennion1983}
carrying out \textit{ab initio} calculations on ferromagnetic systems.

\begin{figure}[h]
\includegraphics[width=8.5cm]{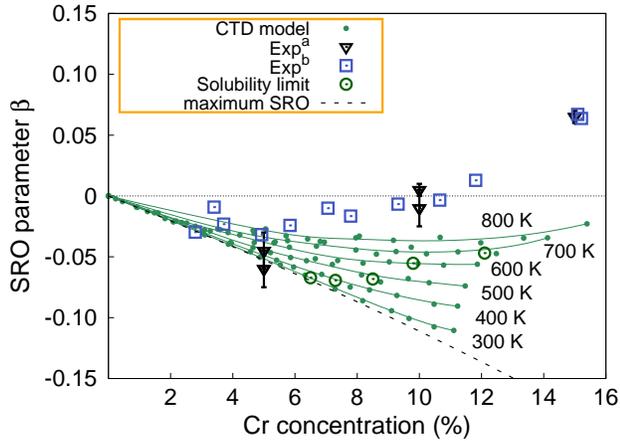}

\caption{\label{fig:sroFeCr}Short range order of the FeCr system as given by the concentration and temperature dependent model (CTD model) with interactions up to $2nn$. The results of the model are compared to available experimental data (blue squares and black triangles)  \cite{Mirebeau1984,Mirebeau2010} performed at 703 K.}
\end{figure}

The analysis of the SRO parameter is of technological importance because
of its implications on the mechanical properties of the alloy. It
is usually defined following the Cowley's notation \cite{Cowley1950,Cowley1950_2}
where the expression for the $i^{th}$ atomic shell of a B atom in
an A-B binary allow is given by: 
\begin{equation}
\alpha_{B}^{(i)}=1-\frac{z_{A}^{(i)}}{z^{(i)}(1-x_{B})}
\end{equation}
 where $z_{A}^{(i)}$ denotes the number of A atoms in the
$i^{th}$ shell from a B atom, $z^{(i)}$ is the total
number of atoms in the $i^{th}$ shell and $x_{B}$ is the global
concentration of B atoms. The value of this parameter will tend to 1 in a segregated alloy and it will be close to 0 
for a random solution. For a system with ordering tendency the value will be negative, with a minimum given by:
\begin{equation}
\alpha_{B}^{(i)}=-\frac{x_B}{1-x_B}
\end{equation}
This latter value indicates the maximum degree of short range order that an alloy can possibly attain.
In the studies by Mirebeau \textit{et al}.
the parameter that is actually measured is specific for BCC structures,
defined for the Fe-Cr system as:
\begin{equation}
\beta=\frac{8\alpha_{Cr}^{(1)}+6\alpha_{Cr}^{(2)}}{14}
\label{eq:sro_bcc}
\end{equation}
To be able to compare to the experimental measurements and to the
recently published data based on the empirical energetic models described
above, we have performed equilibrium Monte Carlo calculations in the semigrand-canonical ensemble \cite{Frenkel2002} 
and measured the parameter
described in eq. \ref{eq:sro_bcc} for different Cr concentrations. 
Results are shown in Fig. \ref{fig:sroFeCr} where the concentration and temperature dependent model was used with interactions up to the $2nn$. We observe that the model slightly overstimates the ordering tendency but captures 
the ordering trend of the alloy for small Cr concentrations. The $\beta$ values tend to $0$ with temperature due 
to entropic effects. The solubility limits are also shown in the figure. Beyond the solubility limit, the SRO parameters are measured in metastable solid solutions which remain homogeneous during the simulation.

We have shown how the model captures the ordering tendency of Cr in the Fe matrix. The $\beta$ parameter becomes negative for low Cr concentrations. Experiments show the same trend, with 
negative values for low Cr concentration and a change in sign at around $11\%$. This inversion of sign observed experimentally is probably due to the presence of a secondary $\alpha '$ phase, as explained by Erhart et al. \cite{Erhart2008-2}. Our results are in very good agreement 
with those presented in \cite{Erhart2008-2} using the CDM semiempirical potential \cite{Caro2005}.

\section{Discussion\label{sec:Discussion}}

The results given above show that the simple
pair interaction model described along the manuscript is able to reproduce
the main features of the experimental phase diagram. In the following
we are going to analyze the matches and the disagreements between
our model and existing models in the literature with special attention to the Cr solubility in Fe, since it controls the precipitation driving force.

\begin{figure}[h]
\includegraphics[width=0.4\textwidth]{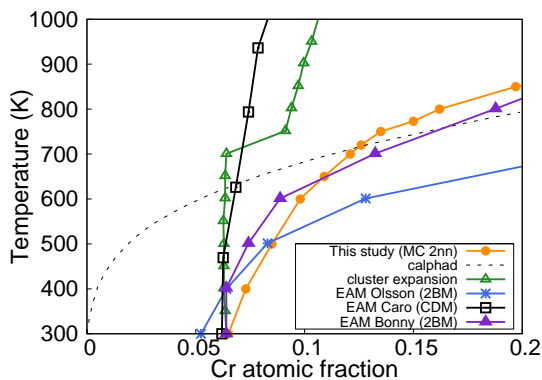}
\caption{\label{fig:pd_lowc} Solubility limits of Cr in iron as given by different models. The Calphad values are taken from Ref. \onlinecite{Andersson1987}. The solubility limit for the CE and for the CDM were presented in Ref. \onlinecite{Bonny2009}. The values for both 2BM potentials have been obtained from Ref. \onlinecite{Bonny2011}.}
\end{figure}

Our model has been fitted to the experimental critical temperature following the CALPHAD
approach by Andersson and Sundman \cite{Andersson1987} (dotted line in Fig. \ref{fig:pd_lowc}) and therefore it reproduces the value given by the regular solution results but with a
solubility limit of Cr in Fe different from zero. The magnetic model proposed by Inden and Sch$\ddot{o}$n \cite{Inden2008} following a cluster variation method was fitted to high temperature values of the experimental phase diagram. It reproduces the magnetic phase transitions with a critical temperature for the miscibility gap of around 880 K. The solubility limits at low temperature tend to $0$ in both sides of the phase diagram. It would be interesting to see the results of this model with the parameters fitted to \textit{ab initio} results. In the magnetic Ising model by Ackland \cite{Ackland2006} the temperature is not in real units and it is hard to compare. However, it reproduces the magnetic transitions in spite of its simplicity and results
in an asymmetric phase diagram.
Concerning the non-magnetic
CE of Lavrentiev \textit{et al}. \cite{Lavrentiev2007, Bonny2009},  the solubility of Cr in Fe at low temperatures reported in that study
 matches the values obtained with our model (see Fig. \ref{fig:pd_lowc}),
even though the curves deviate for temperatures above 400 K, with the solubility
predicted by our approach larger than the one given by the CE. This CE development does not take into account the vibrational or magnetic entropy which results in low solubility at high temperatures.

The semi-empirical interatomic
potentials existing in the literature and described in Sec. \ref{sec: Introduction}
have not been
fitted to the phase diagram itself, but only to the enthalpy of mixing
at 0 K.
The original 2BM predicts a symmetric mixing enthalpy, with two changes of
sign at low and high Cr concentration implying a non-zero Fe concentration
in Cr at low temperatures. The vibrational entropy is found to be very high what 
implies a decrease in the critical temperature to around 750 K (EAM Olsson in Fig. \ref{fig:pd_lowc}). For the new version of the potential, 
the mixing enthalpy is non-symmetric, following the DFT results by Olsson \textit{et al} \cite{Olsson2007} using 
SQS structures. Therefore, the solubility of Fe in Cr is closer to the experimental values. 
The vibrational entropy is lower in this case which increases the critical temperature to a value close to 1100 K (EAM Bonny in Fig.  \ref{fig:pd_lowc}).
The CDM is fitted to the enthalpy of mixing of the alloy as given by exact muffin-tin orbitals theory within the coherent potential approximation (EMTO-CPA) calculations \cite{Olsson2003}. 
The maximum value is, in this case, higher than using SQS-PAW structures. This effect, added to the fact that the vibrational entropy is lower, 
results in a critical temperature above the experimental melting temperature (EAM Caro in Fig. \ref{fig:pd_lowc}). These models have been fitted to 0 K enthalpy of mixing curves. Both Fe and Cr undergo a magnetic transition at high temperatures (around 1043 K for Fe and 312 K for Cr). This means that calculations beyond the magnetic critical temperature of the alloy are out of their scope. 
In a kinetic calculation in a rigid lattice \cite{Bonny2009-3, Pareige2011} using these potentials directly, the vibrational contribution to the entropy is not taken into account, which modifies the solubility limits and therefore the chemical composition in each phase. 
More specifically, the experimental solubility limit at 773 K in the Fe rich side of the phase diagram is about 14-15\% \cite{Novy2009, Xiong2010, Bonny2009}. On the other hand, the original 2BM potential (according to the latest reported values \cite{Bonny2011}) gives no miscibility gap at 773 K, while, if the data shown in Ref. \onlinecite{Bonny2009} is still valid, the solubility limit at the same temperature without the vibrational contribution is around 8\%. The new version of the potential gives a solubility about 20\% and it seems more suitable for these kind of kinetic calculations because its vibrational entropy is lower. However, there is no information about the values for the solubility without taking into consideration the vibrational entropy. For the CDM potential the solubility at this temperature is already too low. Its vibrational entropy contribution is small and, therefore, the values for the solubility limits in a rigid lattice model will not be strongly modified. Although, the variation will persist and care should be taken if kinetic simulations on a rigid lattice are to be performed. This discrepancy between the relaxed models versus the rigid lattice approximations will result in the wrong thermodynamic forces as taken into account in the kinetic calculation. In the studies on precipitation kinetics published in Refs. \onlinecite{Bonny2009-2, Bonny2009-3, Pareige2009, Pareige2011} nucleation starts around 10\%, in disagreement with experiments. It is worth noting that describing this concentration region accurately is important  for industrial applications. 

Our model avoids such a drawback which makes it more useful for kinetic calculations. 
In our model, the pair interactions depend on the temperature
taking in this way into account the magnetic and vibrational contributions
to the entropy. This approximation will not be able to reproduce the magnetic transitions either for the pure elements or the alloy. Even though the ferromagnetic-paramagnetic transition in Fe is not linear with the temperature and neither it is in Cr or the alloy, the simple model described is able to reproduce the experimental phase diagram by adding just one extra degree of freedom. This extra degree of freedom does not affect the computational performance and makes it suitable for large-scale calculations.

\section{Conclusion\label{sec: Conclusions}}

We propose in this article a rigid lattice model based on concentration
and temperature dependent pair interactions to describe the thermodynamics
of Fe-Cr system in the whole concentration range. It is fitted to
both \textit{ab initio} calculations of the enthalpy of mixing at
0 K and to the experimental critical temperature. Only the cross terms
of the atomic pair interactions depend on both the local concentration
and the temperature, while the self interaction terms are fitted to
the respective cohesive energies. In order to check the sensitivity of energetic values in FeCr alloys against DFT implementations with different approximations, and to choose the most accurate values
of the enthalpy of mixing, we have performed a set of first principle
calculations. We carried out the calculations from two different kind
of approaches. The first one was the norm-conserving pseudopotential
approach as implemented in the efficient SIESTA-NC code and the second the more robust projector-augmented
wave as implemented in the PWSCF code. Both approaches give the same qualitative trend, but different quantitative mixing energies at 0 K. The resulting models are similar, although the SIESTA values overestimate slightly the Cr solubility in Fe.

Although its simplicity and even though it does not explicitly consider the magnetic degrees of freedom, this approach captures
the main features of the Fe-Cr thermodynamics: thanks to the concentration
dependence of the pair interactions, it reproduces the transition
between the ordering and demixing tendency, and the trend in the
short-range order parameter when the Cr content increases. The magnetic
and non-configurational entropic contributions are taken into account
by a linear temperature dependence of the pair interactions. The resulting
phase diagram is in very good qualitative and quantitative agreement
with the experimental results. Finally, the model remains simple enough
to be used in Monte Carlo simulations of the solid solution decomposition
kinetics (preliminary results can be found in Ref.\,\onlinecite{Martinez2011}).

\acknowledgments
The authors gratefully acknowledge E. Clouet for useful discussions. This research has received partial funding from the European Atomic Energy Community's 7th Framework Programme (FP7/2007-2011), under grant agreement number 212175 (GetMat project). Part of this work was performed using HPC ressources from GENCI-CINES (Grant 2011-x2011096020). E. Martinez wants to thank the Spanish
Ministry of Science and Innovation, subprogram Juan de la Cierva and the Energy Frontier Research Center, Center for Materials at Irradiation and Mechanical Extremes at Los Alamos National Laboratory (DOE-BES) for partial funding. 


%

\end{document}